\documentclass[conference]{IEEEtran}
\IEEEoverridecommandlockouts
\usepackage{cite}
\usepackage{amsmath,amssymb,amsfonts}
\usepackage{algorithmic}
\usepackage{graphicx}
\usepackage{textcomp}
\usepackage{xcolor}
\usepackage{algorithm}
\usepackage{algorithmic}

\usepackage{amsmath,amsfonts,bm}

\def\modelname{Dictionary LISTA}
\def\modelacro{DLISTA}

\def\Figref#1{Figure~\ref{#1}}

\def\secref#1{section~\ref{#1}}
\def\Secref#1{Section~\ref{#1}}

\def\eqref#1{equation~\ref{#1}}

\def\1{\bm{1}}

\def\vg{{\bm{g}}}
\def\vh{{\bm{h}}}

\def\vp{{\bm{p}}}

\def\vr{{\bm{r}}}

\def\vx{{\bm{x}}}
\def\vy{{\bm{y}}}
\def\vz{{\bm{z}}}

\def\mD{{\bm{D}}}

\def\mH{{\bm{H}}}

\def\mP{{\bm{P}}}

\def\mU{{\bm{U}}}
\def\mV{{\bm{V}}}

\def\mZ{{\bm{Z}}}

\def\mPhi{{\bm{\Phi}}}
\def\mPsi{{\bm{\Psi}}}

\def\mSigma{{\bm{\Sigma}}}
\def\mOmega{{\bm{\Omega}}}

\DeclareMathAlphabet{\mathsfit}{\encodingdefault}{\sfdefault}{m}{sl}
\SetMathAlphabet{\mathsfit}{bold}{\encodingdefault}{\sfdefault}{bx}{n}

\DeclareMathOperator*{\argmax}{arg\,max}
\DeclareMathOperator*{\argmin}{arg\,min}

\newcommand*{\norm}[1]{\left\|#1\right\|}
\newcommand*{\card}[1]{\left|#1\right|}

\usepackage{todonotes}

\def\eqdef{{:=}}

\usepackage{tabularx,booktabs}
\usepackage[printonlyused]{acronym}
\acrodef{AMP}{Approximate message passing}
\acrodef{AoA}{Angle of arrival}
\acrodef{AoD}{Angle of departure}

\acrodef{CDL}{clustered delay line}
\acrodef{CSI}{channel state information}
\acrodef{CIR}{channel impulse response}
\acrodef{DLISTA}{Dictionary LISTA}
\acrodef{IHTA}{iterative hard thresholding algorithm}
\acrodef{ISTA}{iterative soft thresholding algorithm}
\acrodef{LISTA}{learning iterative soft thresholding algorithm}

\acrodef{MMSE}{Minimum Mean Square Error}
\acrodef{MNSE}{Mean Normalized Square Error}
\acrodef{NSE}{Normalized Square Error}
\acrodef{NN}{Neural Network}
\acrodef{OMP}{Orthogonal Matching Pursuit}
\acrodef{PCA}{Principle Component Analysis}

\newcommand{\fv}[1]{{#1}}

\def\BibTeX{{\rm B\kern-.05em{\sc i\kern-.025em b}\kern-.08em
    T\kern-.1667em\lower.7ex\hbox{E}\kern-.125emX}}
\IEEEoverridecommandlockouts\IEEEpubid{\makebox[\columnwidth]{ 978-1-6654-3540-6/22~\copyright~2022\ IEEE \hfill} \hspace{\columnsep}\makebox[\columnwidth]{ }}
\begin{document}
\title{
Beyond Codebook-Based Analog Beamforming at mmWave: Compressed Sensing and\\ Machine Learning Methods

}
\author{
    \IEEEauthorblockN{Hamed Pezeshki\IEEEauthorrefmark{1}, Fabio Valerio Massoli\IEEEauthorrefmark{2}, Arash Behboodi\IEEEauthorrefmark{2}, Taesang Yoo\IEEEauthorrefmark{1}, 
    }
        \IEEEauthorblockN{
Arumugam Kannan\IEEEauthorrefmark{1}, Mahmoud Taherzadeh Boroujeni\IEEEauthorrefmark{1}, Qiaoyu Li\IEEEauthorrefmark{3}, Tao Luo\IEEEauthorrefmark{1}, Joseph B. Soriaga\IEEEauthorrefmark{1},
    }
    \IEEEauthorblockA{\IEEEauthorrefmark{1}Qualcomm Technologies, Inc. 
    \IEEEauthorrefmark{2}Qualcomm Technologies Netherlands B.V.,\\
    \IEEEauthorrefmark{3}Qualcomm Wireless Communication Technologies 
    }
}
\maketitle
\vspace{-3mm}
\begin{abstract}
Analog beamforming is the predominant approach for millimeter wave (mmWave) communication given its favorable characteristics for limited-resource devices. In this work, we aim at reducing the spectral efficiency gap between analog and digital beamforming methods. We propose a method for refined beam selection based on the estimated raw channel. The channel estimation, an underdetermined problem, is solved using compressed sensing (CS) methods leveraging angular domain sparsity of the channel. To reduce the complexity of CS methods, we propose dictionary learning iterative soft-thresholding algorithm, which jointly learns the sparsifying dictionary and signal reconstruction. We evaluate the proposed method on a realistic mmWave setup and show considerable performance improvement with respect to code-book based analog beamforming approaches. 
\end{abstract}

\section{Introduction}

The de-facto method for communication at mmWave frequencies in existing implementations is analog beamforming using predefined codebooks at transmitters and receivers. While analog beamforming makes mmWave communications practically realizable by allowing hardware architectures with limited RF-baseband chains, the purpose of this paper is to explore what could be done beyond the said approach by inferring additional information about the underlying raw channel using compressed sensing and machine learning tools. 

Compressed Sensing (CS) methods have been widely used to solve the channel estimation problem in context of \fv{millimiter-wave (mmWave)} and massive MIMO~\cite{rodriguez-fernandez_frequency-domain_2018,venugopal_channel_2017,wang_beam_2019,ma_high-resolution_2020}. Such an estimation problem is underdetermined since, for example, analog beamforming provides lower dimensional, e.g., 2 by 2, observation of a higher dimensional channel, e.g., 8 by 64. 
Classically, CS methods leverage the sparsity of the channel in the angular domain. In such a scenario, the basis vectors with respect to which the channel can be expressed as a sparse vector is defined as the outer product of antenna responses on the transmit and receive side. We typically refer to such vectors as atoms of the sparsifying dictionary. 
The channel estimation problem can therefore be cast as a Sparse Recovery Problem (SRP).
There are several methods in the CS literature to solve the SRP with some notable examples of iterative methods such as \ac{OMP} \cite{pati_orthogonal_1993,davis_adaptive_1994}, \ac{IHTA} \cite{blumensath_iterative_2009}, \ac{ISTA} \cite{daubechies2004iterative}, and \ac{AMP} \cite{donoho_message-passing_2009}. Two restricting factors of classical CS methods for deployment on limited resource devices are dimension of the sparsifying dictionary and the number of iterations.  The hyperparameters of these algorithms, e.g., number of iterations, can be tuned to obtain better performance-complexity point.
Machine learning methods can be used to reduce either the number of iterations or the number of atoms.  
Learning Iterative
Soft Thresholding Algorithm (LISTA) and its variations~\cite{behrens_neurally_2021,gregor_learning_2010,liu_alista_2019,chen_hyperparameter_lista_2021} focus on solving the SRP for fixed sensing matrix and sparsifying dictionary by unfolding iterative recovery algorithms as neural network layers and learning parameters from the data. For more general cases, follow-up works try to either learn a dictionary suitable for reconstruction tasks or adapting the parameters to new sensing matrices~\cite{aberdam_ada_lista_2021,schnoor_generalization_2022}.
 
In this work, we consider mmWave model in its generality with planar dual polarization antenna arrays in 3D space. This setup is more challenging as the dimensionality of the problem blows up, and the SRP becomes significantly more complex. 
In our work, we show how the information about raw channel can be used to improve spectral efficiency by tailoring the analog beam selection to the underlying raw channel, as opposed to relying on a pre-defined codebook. We consider two approaches to obtain the underlying channel from beamformed measurements. First, we use classical compressed-sensing methods like \ac{OMP}.
{To reduce the complexity of \ac{OMP}, we formulate Dictionary LISTA (DLISTA), a machine learning model based on an unrolled version of \ac{ISTA} in which each iteration is interpreted as the layer of a neural network with learnable parameters. \modelacro{} jointly learns the sparsifying dictionary and the reconstruction algorithm. It can considerably reduce the number of iterations as well as the dictionary dimensions.} 
Differently than \cite{schnoor_generalization_2022}, our method focuses on variable sensing matrices, which is more challenging to train. The rest of the paper is organized as follows. In Section \ref{sec:definitions}, the main problem setup is presented. We discuss the classical compressed sensing approach with beam selection process in Section \ref{sec:beam_selection}. We describe \modelacro{} in Section \ref{sec:dict_learning}. The experiments are presented in Section \ref{sec:experiments}.


\section{System Model and Problem Statement}
\label{sec:definitions}
\begin{figure*}
    \centering
    \includegraphics[width=.9\textwidth]{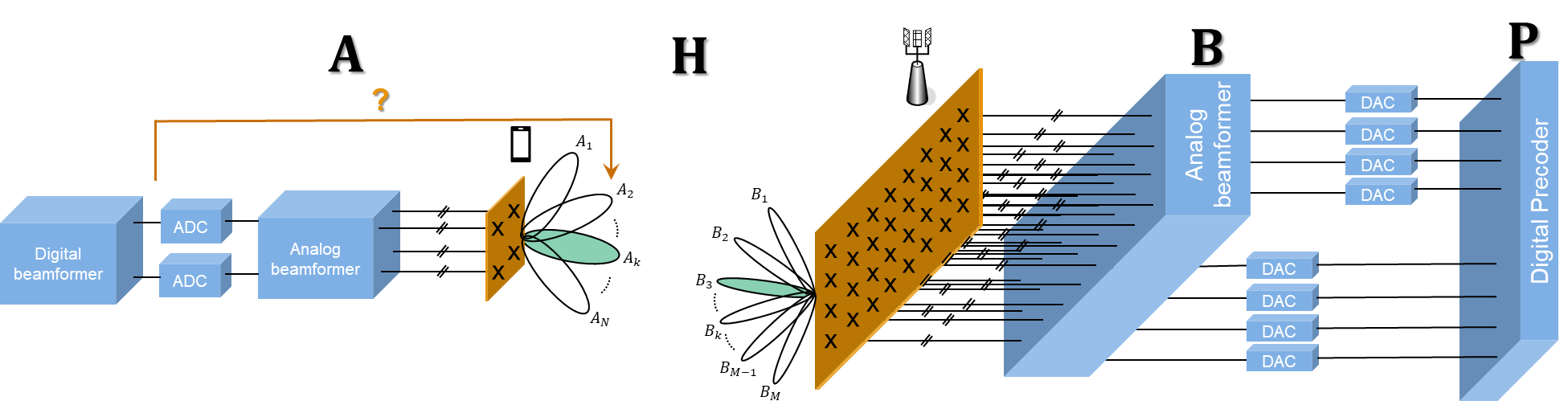}
    \caption{System Model}
    \label{fig:system_model}
\end{figure*}
Consider a mmWave analog beamforming scenario as in Figure \ref{fig:system_model}. To enable communication in mmwave frequencies, transmitters and receivers rely on pre-defined analog beamforming codebooks in existing implementations. In order to find the best beam(s) to communicate with, from the pre-defined codebook, transmitters and receivers go through a beam management procedure. Let us assume downlink transmission in a cellular setting and further assume that the receiver (user equipment, UE) and the transmitter (gNB) have $N_\text{UEant}$ and $N_\text{NBant}$ dual-polarization antennas, respectively. The number of utilized RF chains at the UE and gNB side are $N_\text{UErf}$ and $N_\text{NBrf}$, respectively. The analog beamforming matrices utilized at the UE and gNB are given respectively by $\mathbf{A}_i [N_\text{UErf}\times N_\text{UEant}]$ and $\mathbf{B}_i [N_\text{NBant}\times N_\text{NBrf}]$, where the index $i$ is chosen from the total number of beams in UE and gNB codebook  $N_\text{UEbeams}$ and $N_\text{NBbeams}$, respectively. The TX- and RX-beamformed channel for the $i$-th beam pair is $\textbf{A}_i\textbf{H}\textbf{B}_i$ whose dimension is $[N_\text{UErf}\times N_\text{NBrf}]$, which is considerably lower than the dimension of the raw channel $\textbf{H} [N_\text{UEant}\times N_\text{NBant}]$. We consider the following problem: {Given multiple of these TX- and RX-beamformed measurements, \textit{what} information can we infer about the underlying raw channel and \textit{how} can we use this information to improve system performance?} This is an under-determined problem, and we use compressed sensing and machine learning tools to address the problem in this paper. The sparsity structure of the channel is revealed using geometric channel model \cite{rodriguez-fernandez_frequency-domain_2018,schniter_channel_2014}. We extend this model to 3D environment where each arrival and departure direction is characterized by angle and zenith values, yielding $(AoA,ZoA,AoD,ZoD)$ quadruples. The channel $\mH_d$ of delay tap $d$ is given by:
\begin{equation}
    \mH_d = \sum_{l=1}^L \alpha_l\vp_R(\phi^{AoA}_l,\phi^{ZoA}_l)\vp_T^*(\phi^{AoD}_l,\phi^{ZoD}_l),
    \label{eq:geometric_channel_model}
\end{equation}
where $\alpha_l$ contains the effect of path loss, lowpass filtering and complex gains, and $\vp_R,\vp_T$ are receive and transmit antenna element response, respectively. Finally, $L$ is the number of channel paths for $d$-th delay tap of the channel. We consider a conventional MIMO-OFDM based transmission over multiple subcarriers. In this paper, we consider DFT codebooks.
\section{Estimation of Raw mmWave Channel}
\label{sec:beam_selection}
We consider two methods for inferring underlying raw channel based on beamformed measurements. The first method is based on classical compressed sensing-based approach, in which we estimate channel \acp{AoA} and \acp{AoD}. We then propose a method about how to use this acquired information to enhance system performance. The second method is a fixed complexity recovery method with lower dimensional dictionary. 
\subsection{Classical compressed sensing approach}
Using the channel model in \eqref{eq:geometric_channel_model}, we can find an approximate sparse representation of the channel $\mH_d$ given by angular quantization
\begin{equation}
\mathbf{H}_d \approx \mathbf{P}_R\mathbf{\Delta}_d^q\mathbf{P}_T^*,
\label{eq:sparse_rep_H}
\end{equation}
where $\mathbf{P}_R [N_\text{UEant}\times[N_\text{UEazi}N_\text{UEelev}]]$ and $\mathbf{P}_T [N_\text{NBant}\times [N_\text{NBazi}N_\text{NBelev}]]$ are the antenna element response matrices, with columns $\vp_R$ and $\vp_T$ computed over 2-D $N_\text{UEazi}\times N_\text{UEelev}$ and $N_\text{NBazi}\times N_\text{NBelev}$ discretized angular grid points  at the receiver and transmitter, respectively. The matrix $\mathbf{\Delta}_d^q [[N_\text{UEazi}N_\text{UEelev}]\times[N_\text{NBazi}N_\text{NBelev}]]$ is a sparse matrix with its non-zero elements representing $(AoA,ZoA,AoD,ZoD)$ quadruples along which the channel has a non-zero gain. Note that the approximation in \eqref{eq:sparse_rep_H} is exact if the angular quadruples are on the chosen grid. The grid size choice is a trade-off between performance and complexity. We can rewrite the channel in vectorized format using properties of the Kronecker product:
\begin{equation}\label{channel}
    \text{vec}(\mathbf{H}_d) \approx \text{vec}(\mathbf{P}_R\mathbf{\Delta}_d^q\mathbf{P}_T^*) = \left[\left(\mathbf{P}_T^*\right)^T\bigotimes \mathbf{P}_R\right]\text{vec}(\mathbf{\Delta}_d^q).
\end{equation}
We denote $\mathbf{\Psi} \eqdef \left(\mathbf{P}_T^*\right)^T\bigotimes \mathbf{P}_R$ as the channel sparsifying dictionary. 
With this quantized channel representation, we consider the $d$-th delay tap of the channel impulse response (CIR) for $i$-th TX- and RX-beamformed channel, $\mathbf{A}_i\mathbf{H}_d\mathbf{B}_i$. We have
\begin{align}
    \textbf{y}_{d,i} &\eqdef \text{vec}(\mathbf{A}_i\mathbf{H}_d\mathbf{B}_i) \nonumber\\
    &= \left[(\mathbf{B}_i)^T\bigotimes \mathbf{A}_i\right]\text{vec}(\mathbf{H}_d) = \mathbf{\Phi}_i\mathbf{\Psi}\text{vec}(\mathbf{\Delta}_d^q)
    \label{eq:sparse_rep_y}
\end{align}
For $M$ beamformed channel measurements, we can stack the measurements and get
\begin{equation}\label{recovery_eq}
    \mathbf{y}_d \eqdef \begin{bmatrix} \mathbf{y}_{d,1}  \\ \mathbf{y}_{d,2} \\ \vdots \\ \mathbf{y}_{d,M}  \end{bmatrix}=\begin{bmatrix} \mathbf{\Phi}_1  \\ \mathbf{\Phi}_2 \\ \vdots \\ \mathbf{\Phi}_M  \end{bmatrix} \mathbf{\Psi} \text{vec}(\mathbf{\Delta}_d^q) = \mathbf{\Phi}\mathbf{\Psi} \text{vec}(\mathbf{\Delta}_d^q).
\end{equation}
The raw channel recovery can be seen as a sparse recovery problem
with the matrix $\mathbf{\Phi}\mathbf{\Psi}$ as the overall sensing matrix. Once we estimate $\text{vec}(\mathbf{\Delta}_d^q)$ based on $\mathbf{y}_d$, we can substitute it in \eqref{channel} to recover the channel. In order to solve \eqref{recovery_eq}, we use orthogonal matching pursuit (OMP) \cite{pati_orthogonal_1993}. The details of the OMP algorithm are depicted in Table \ref{alg:OMP}.

\begin{figure}
    \centering
    \includegraphics[width=0.49\textwidth]{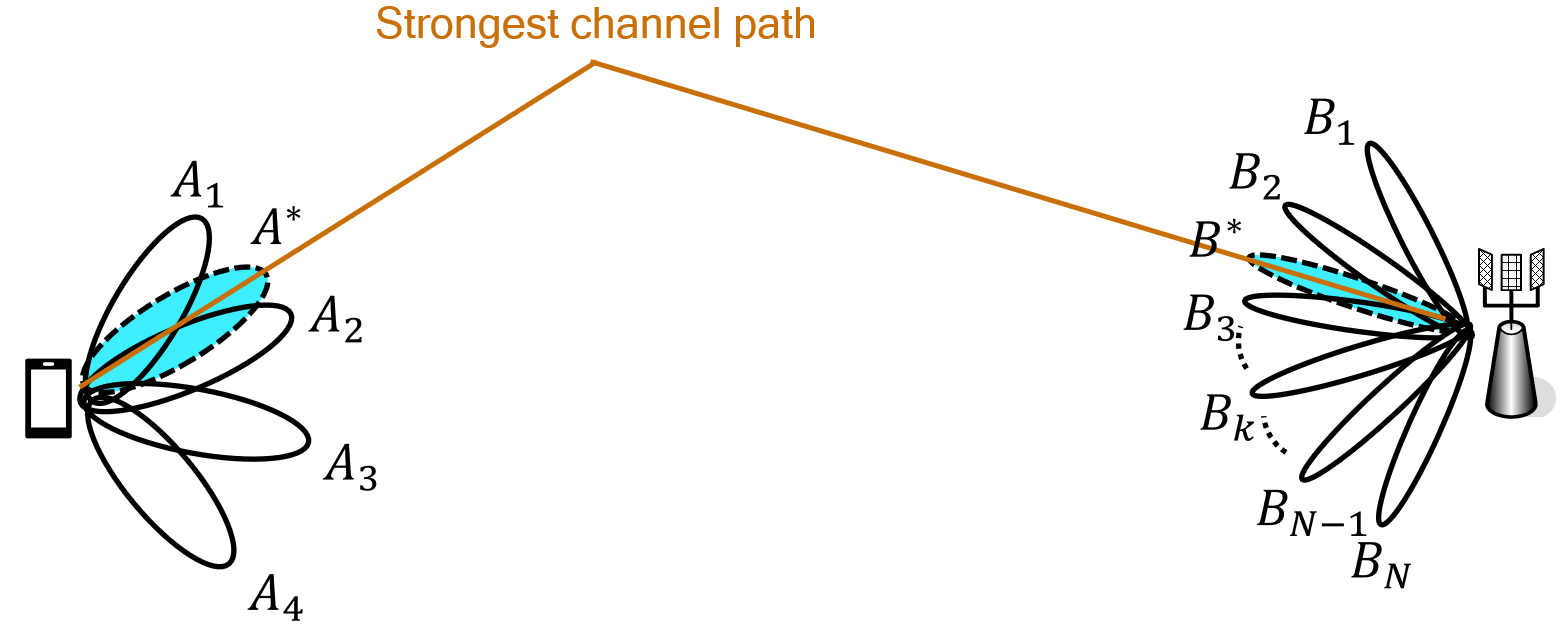}
    \caption{Illustration of non-codebook-based beams tailored to channel AoA/AoD characteristics}
    \label{fig:custom_beam}
\end{figure}

The OMP algorithm is being run per-tap. Leveraging the sparsity of the mmWave channel in the tap domain, the algorithm only needs to be run for a few dominant taps, which is what we do in this work. The OMP algorithm provides the most likely  $(AoA, AoD, ZoA, ZoD)$ at each step. Then through an iterative process, the contribution of the identified angles is subtracted from the observation vector and residual is computed. The algorithm is iterated up to a point where a certain criteria is met (e.g., MSE of the residual is less than a threshold) or for a fixed number of iterations.
Via the OMP algorithm, we can estimate the angular information associated with the strongest channel paths. Using this information about the underlying raw channel, we can create a \textit{custom} beam (not a beam from transmitter or receiver codebook) at the transmitter or receiver with their beam pointing angles directed along the estimated AoD-AoA of the strongest channel cluster, as depicted in Fig. \ref{fig:custom_beam}. In Section \ref{sec:experiments}, we illustrate how using these custom beams can lead to spectral efficiency improvements through system-level simulations.
\begin{algorithm}[t]
\caption{OMP for mmWave channel estimation}\label{alg:OMP}
\begin{algorithmic} 
\REQUIRE Observation $\vy_d$, sensing matrix $\mPhi$, dictionary $\mPsi$
\STATE Initialize $\hat{\mP_T}=\emptyset, \hat{\mP_R}=\emptyset, \vy'_d=\vy_d$
\REPEAT
\STATE $\hat{l}=\argmax_{i}\card{\left((\Phi\Psi)^*\vy'_d\right)_i}$
\STATE Extract $(AoA_{\hat{l}},ZoA_{\hat{l}},AoD_{\hat{l}},ZoD_{\hat{l}})$
\STATE $\hat{\mP_T} \gets \left[\hat{\mP_T} \quad \vp_T(AoD_{\hat{l}},ZoD_{\hat{l}})\right]$
\STATE $\hat{\mP_R} \gets \left[\hat{\mP_R} \quad \vp_R(AoA_{\hat{l}},ZoA_{\hat{l}})\right]$
\STATE $\hat{\Psi} \gets \left[\hat{\mP_T}^T\bigotimes \hat{\mP_R}\right]$
\STATE $\hat{\vx_d} \gets (\Phi\hat{\Psi})^*\vy_d$
\STATE $\vy'_d \gets \vy_d-\Phi\hat{\Psi}\hat{\vx_d}$
\UNTIL Termination criteria is met
\RETURN  $\hat{\vx_d},\hat{\Psi}$
\end{algorithmic}
\end{algorithm}
\vspace{-3mm}
\subsection{Dictionary learning for channel estimation}
Classical compressed sensing approaches faces two challenges. First, the sparsifying dictionary
$\mathbf{\Psi} = \left(\mathbf{P}_T^*\right)^T\bigotimes \mathbf{P}_R$ can be very high-dimensional. As an example, defining a grid of 1-degree resolution over angular quadruples would lead to $180^2\times 360^2$ atoms in the dictionary. The other challenge in the number of iterations needed for convergence. One solution to reduce the complexity of classical methods is to learn a lower-dimensional dictionary. This is possible for a specific environment, because the channel instances possess additional structures reflecting spatial characteristics of the environment. These structures are more difficult to capture analytically, and dictionary learning aims at representing this structure as sparse representation in terms of some learned basis elements. We can learn the dictionary from a dataset of channel traces. Suppose that the channel traces are given as $\left(\mH_i\right)_{i\in[m]}$. We construct the data matrix from the vectorized traces $\mH = [\text{vec}(\mH_1) \dots \text{vec}(\mH_m)]$. The dictionary learning problem can be stated as:
\begin{align}
\min_{\mD,\mZ} & \|\mH-\mD\mZ\|_F^2 \nonumber\\
&\text{ subject to: } \mD\in\mOmega, \mZ=[\mZ_1\dots\mZ_m],  \mZ_i\in\mSigma_s,
\end{align}
where $\mSigma_s$ is the set of $s-$sparse vectors, and $\mOmega$ is a set of matrices of choice, for example, orthogonal matrices. For the rest, we choose the set  $\mOmega$ as the set of orthogonal matrices. We use two methods to solve the above problem, namely sparse PCA (SPCA) \cite{ma_sparse_2013} and kSVD \cite{aharon2006k}. We formulate an iterative hard-thresholding procedure for sparse PCA based on two steps, namely, dictionary update and sparse representation update. The goal was to have direct control over the sparsity order using hard-thresholding operator. We report the details for the SPCA in Algorithm~\ref{alg:spca}.

\begin{algorithm}[t]
\caption{Iterative Hard Thresholding Sparse PCA}\label{alg:spca}
\begin{algorithmic} 
\REQUIRE Data matrix $\mH = [\text{vec}(\mH_1) \dots \text{vec}(\mH_m)]$, number of iterations ${T}$, threshold function $\eta_\theta$
\STATE $Z^{(0)} \sim \mathcal{N}(\mathbf{0}, \mathbf{I})$
\FOR{$i \gets 1$ to $\mathcal{T}$}
\STATE $\mD^{(i)} \gets\argmin_{\mD} \|\mH-\mD\mZ^{(i-1)}\|_F^2$, which is $\mD^{(i)}=\mU\mV^*$ from SVD of $\mH(\mZ^{(i-1)})^* = \mU \mSigma \mV^*$
\STATE $\mZ^{(i)} \gets H_s [(\mD^{(i)})^*\mH]$ where $H_s(\cdot)$ is the hard-thresholding operator applied column-wise to keep top $s$ values.
\ENDFOR
\RETURN Sparsifying dictionary $\mD^{(T)}$
\end{algorithmic}
\end{algorithm}

In Algorithm~\ref{alg:spca}, we set the number of iterations to 30 and employ the hard threshold function, selecting the highest 10\% of the input's entries. Concerning the kSVD, we use the formulation reported in~\cite{aharon2006k,rubinstein2008efficient}. Being more computationally demanding compared to SPCA, for kSVD we set the number of iterations to one. However, as pointed out in~\cite{rubinstein2008efficient}, one iteration is enough to evaluate a set of sparsifying vectors retaining enough structure to reconstruct the original signal.

\section{\modelname{} for channel estimation}
\label{sec:dict_learning}
In order to exploit the full power of ML techniques, we formulate \modelacro{}: a \ac{NN} model based on the ISTA~\cite{daubechies2004iterative,schnoor_generalization_2022} iterative method to solve the SRP. 
Our proposed method, differentiates from the classical and other ML approaches proposed in the literature~\cite{gregor_learning_2010,liu_alista_2019,behrens_neurally_2021,aberdam_ada_lista_2021,chen_hyperparameter_lista_2021} mainly for two aspects. On the one hand, it is not limited to a fix measurement matrix, and on the other hand, it can learn the appropriate sparsifying dictionary jointly while solving the recovery problem.

\begin{figure}[t]
    \centering
    \includegraphics[width=\linewidth]{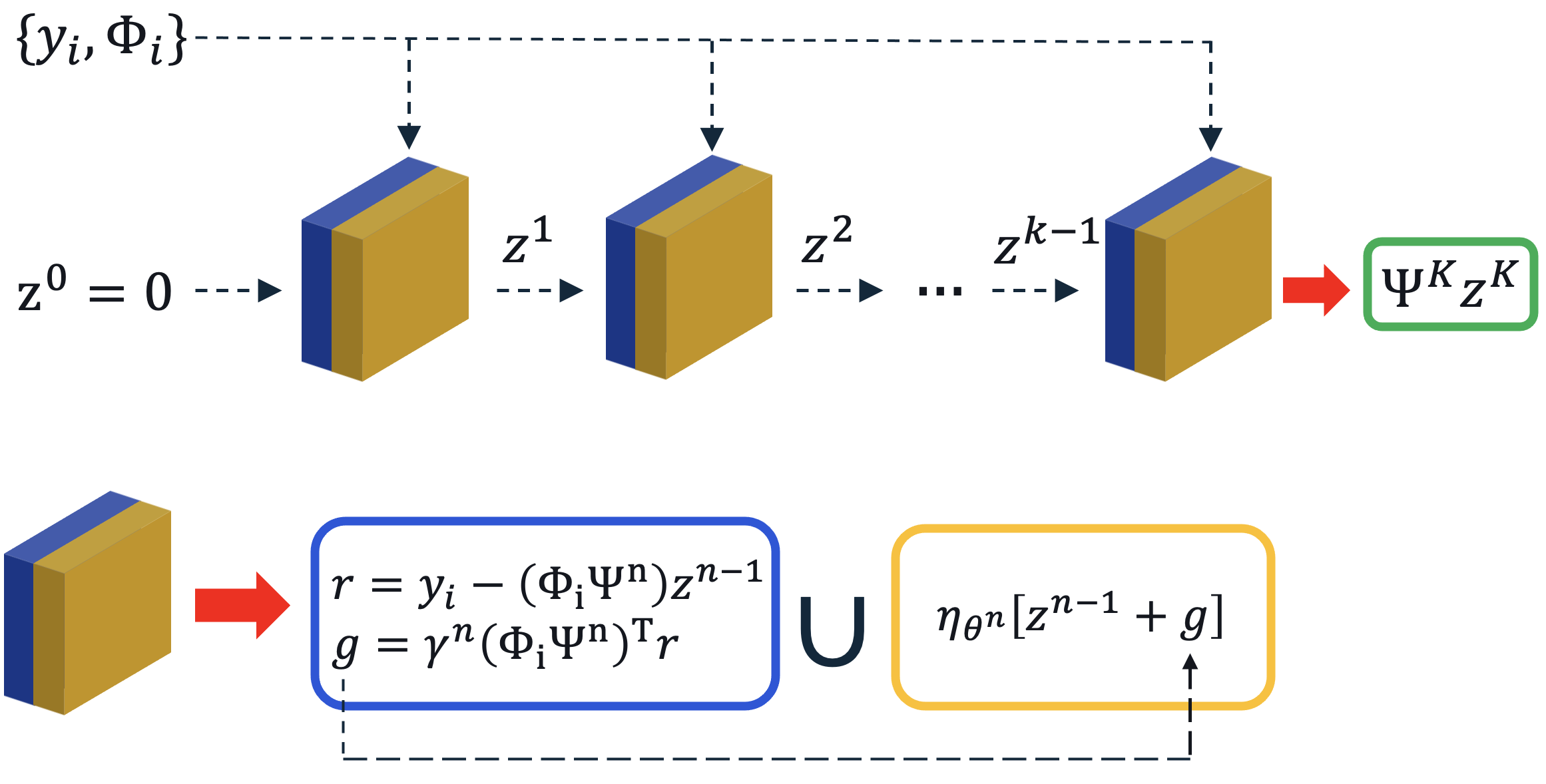}
    \caption{Top: Schematic representation of the \modelacro{} architecture. Bottom: operations performed at the $n$-th iteration: update of the sparse vector (blue block) and soft-thresholding (dark yellow block).}
    \label{fig:dlista}
\vspace{-5mm}
\end{figure}

The use of a varying sensing matrix is especially relevant in the context of mmWave communication, since UE and gNB can select different set of beams for measurements in each round. To the best of our knowledge, the only other method capable of handling such a scenario is proposed in~\cite{aberdam_ada_lista_2021}. However, also in that case the model requires a priori knowledge about the sparsifying set of basis vectors for the signal to be recovered. Moreover, it requires for the dictionary to be fixed. Instead, \modelacro{} can either train its own dictionary from a randomly initialized one or accept, and eventually "fine tune'', a pre-computed dictionary. We report in \Figref{fig:dlista} a schematic view of the \modelacro{} model. 
\modelacro{} is based on unrolling the computation graph of iterative soft-thresholding algorithm (ISTA). ISTA is the proximal gradient descent solution for LASSO defined for an observation $\vy_i$ and sensing matrix $\Phi_i$ as :
\begin{equation}
    \argmin_{\vz}\norm{\vy_i-\mPhi_i\mPsi\vz}_F^2+\lambda\norm{\vz}_1.
\end{equation}
The main step of the algorithm at iteration $k$ is given by
\begin{equation}
    \vz^{k}=\eta_{\theta^k}\left(\vz^{k-1}+\gamma^k(\vy_i-\mPhi_i\mPsi\vz^{k-1}) \right),
\end{equation}
where $\eta_\theta(\cdot)$ is the shrinkage operator with parameter $\theta$ applied entry-wise and defined as $\eta_\theta(x)=ReLU(x-\theta)-ReLU(-x-\theta)$\footnote{$ReLU(\cdot)$ is the rectified linear function.}. \modelacro{} specifies a map $f_\Theta: (\mathbb{C}^n, \mathbb{C}^{n \times m}) \rightarrow \mathbb{C}^r$ that, given a pair $\vy_i, \mPhi_i \sim \mathcal{D}$ containing the observation at the UE side and the sensing matrix for the $i$-th data sample, respectively, estimates the channel matrix. The function $f_\Theta$ is defined by a fixed number of  ISTA iteration layers and parametrized by a set of learnable parameters $\Theta=\{\gamma^k, \theta^k, \mPsi^k \}_{k=0}^{K-1} \cup \{\mPsi^K\}$ employed in the iterative procedure. The last learnable matrix $\mPsi^K$ is used to estimate the channel from its sparse representation. Note that $\mPsi^k$ can be fixed to $\mPsi$ and shared among all the layers. We report in Algorithm~\ref{alg:dl_lista} the pseudocode to describe the iterative procedure as well as the training loop for the model's parameters.

\begin{algorithm}
\caption{\modelacro{} training}\label{alg:dl_lista}
\begin{algorithmic} 
\REQUIRE $\mathcal{D}=\{\vy_i, \mPhi_i, \vh_i\}_{i=1}^N$; number of iterations $\mathcal{K}$; $\Theta=\{\gamma^k, \theta^k, \mPsi^k\}_{k=0}^{\mathcal{K}-1} \cup \{\mPsi^{\mathcal{K}}\}$; objective $\mathcal{L}$
\FOR{$i \gets 0$ to $\mathcal{N}-1$}
\STATE $\vz^{k=0}_i \gets 0$
\FOR{$k \gets 0$ to $\mathcal{K}-1$}
\STATE $\vr \gets \vy_i - \mPhi_i \mPsi^k \vz^{k}$
\STATE $\vg \gets \gamma^k\Big(\mPhi_i \mPsi^k \Big)^*\cdot \vr$
\STATE $\vz^{k+1}_i \gets \eta_{\theta^k}\Big(\vz^k_i + \vg\Big)$
\ENDFOR
\STATE $\hat{\vh}_i=\mPsi^\mathcal{K} \vz^\mathcal{K}_i$ 
\STATE $\mathcal{L}=NMSE(\vh_i, \hat{\vh}_i)$
\ENDFOR
\RETURN $\Theta=\{\gamma^k, \theta^k, \mPsi^k\}_{k=0}^{\mathcal{K}-1} \cup \{\mPsi^{\mathcal{K}}\}$
\end{algorithmic}
\end{algorithm}

We can break down the iteration procedure as follows. First, the sparse vector is initialized to contained only zero elements, $z^0 = 0$. Next, \modelacro{} first evaluates the residual among the current observations and the predicted ones based on the estimated sparse vector. Such a residual is then projected back onto the sparse vector space in order to evaluate its update. It is fundamental to not confuse such a step with the backpropagation procedure we use in order to update the value of the learnable parameters. Subsequently, the sparse vector is updated, and the soft threshold function is applied to it. Finally, given the estimated channel, the objective function is evaluated in order to compute the gradients needed to train the model. 
As objective, we adopt the Normalized Mean Squared Error (NMSE) defined as:
\begin{align}\label{eq:nmse_obj}
    \mathcal{L}(\vh, \hat{\vh}) &= 10 \cdot \mathrm{log}\ \mu,\quad \mu =  \frac{1}{N} \sum_i \frac{\Vert \vh_i - \hat{\vh}_i \Vert_2^2}{\Vert \vh_i \Vert_2^2}
\end{align}

After the training, the model is fixed and used for inference as a simple feed forward network. The complexity reduction comes with the price of loosing interpretability of sparse representation. Unlike OMP based methods, the sparse representation and the active elements of learned dictionary do not immediately carry information about angular domain. Therefore, one cannot direct the beam to the dominant cluster. In this case, we select the beam by exhaustive search over an oversampled codebook knowing the raw channel. As last remark, note that the learned dictionary provides a way for sparse representation of the channel, which can be used for other tasks like channel compression and feedback.

\section{Experiments}
\label{sec:experiments}
\begin{figure}
    \centering
    \includegraphics[width=0.49\textwidth]{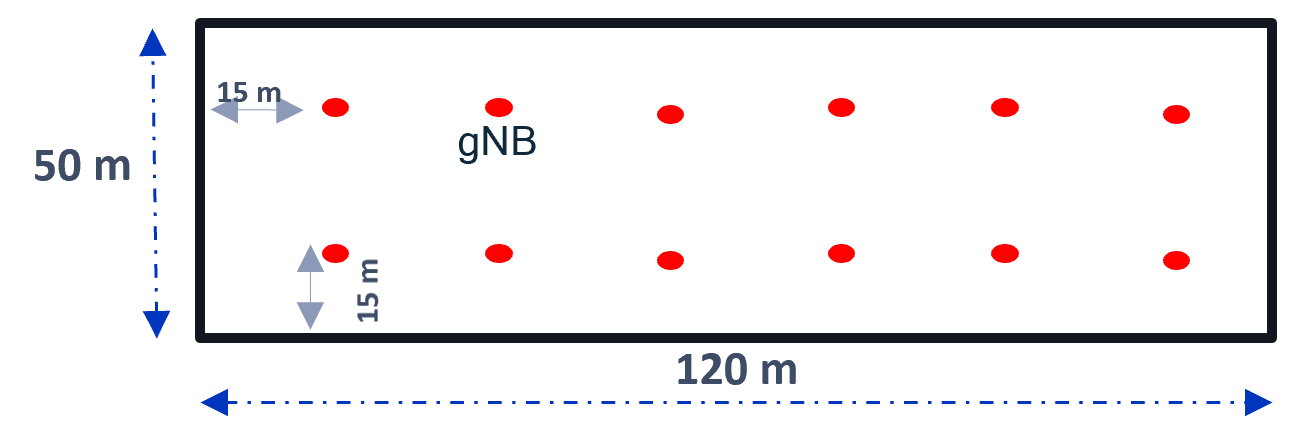}
    \caption{Indoor Hotspot deployment}
    \label{fig:InH_model}
\vspace{-5mm}
\end{figure}

\subsection{Dataset Description and \modelacro{} Training}\label{sec:dataset}
We have used indoor hotspot channel model as specified in 38.901 3GPP document \cite{3gpp_38901_2020}. It represents an indoor
environment with multiple gNBs, see Figure \ref{fig:InH_model}. The simulation parameters are provided in Table \ref{tab:sim_param}. 
\begin{table}[h]
    \centering
    \begin{tabular}{c|c}
    Parameters & Value\\
    \hline
    \hline
        Carrier Freq. & 28 GHz \\
        gNB antenna  & 32 antennas, dual polarization \\
        UE antenna  & 4 antennas, dual polarization \\
        gNB antenna gain  & 5dBi \\
        gNB TX power  & 23 dBm for 100MHz bandwidth \\
        UE RX noise figure  & 13 dB \\
        Subcarrier spacing  & 120 KHz \\
        Number of tones  & 4096 \\
    \hline
    \end{tabular}
    \caption{Simulation Parameters}
    \label{tab:sim_param}
\vspace{-5mm}
\end{table}
In order to learn the parameters of our models, we employ a supervised procedure. The dataset comprises of 120 UEs samples. The communication happens over 4096 tones. Given the fact that the channel is sparse in the tap domain, the channel for each UE has only a few dominant taps. This amounts to an average of three taps per UE. Thus, in total we have $\sim360$ data samples. 
The five highest Reference Signal Received Power (RSRP) beamformed measurements are used for channel estimation. Thus, for each data sample in the dataset, we considered a $5\times 1$ measurement $\vy_i$ with corresponding sensing matrix $\mPhi$. 
In such a scenario, $\mPhi$ represents an underdetermined system of equations having more columns than rows, i.e., more degree of freedom than constraints, thus allowing us to recast the channel estimation problem as a sparse recovery problem that we solve with \modelacro{}. To monitor and test the performance of the model,  we split the dataset into train, validation, and test sets with the following ratios: 0.75, 0.08, 0.17. The use of a validation set is fundamental, for example, to immediately spot training pitfalls such as overfit. Indeed, we notice that due to the low cardinality of the training set, the NN was prone to overfit, thus reporting very poor performance on the test dataset. Thus, we imposed strong regularization through the optimizers and employed early stopping. To train the model we use the Adam
optimizer with a different initial value for the learning rate for the different parameters, and we employ mini-batches. Concerning the parameters, we train the model with shared and non-shared weights and compared the performance. As a general remark, we empirically notice that better performance generally corresponds to NN with no weights sharing. Regarding the sparsifying dictionary, $\mPsi$, we adopt two different strategies for initialization: sparse PCA or random sampling. Concerning the dictionaries, we train the model both with fixed and non-fixed $\mPsi$ matrices. In the first case, the $\mPsi$ matrices are not learnable parameters while, in the second case, we let \modelacro{} train the dictionaries (in this case the sparse PCA and random sampling are used as initialization procedures for the $\mPsi$s analogously as other methods in the ML literature are used to initialize weights and biases for the various layers of a NN). As mentioned in~\Secref{sec:dict_learning}, we use the NMSE as an objective to train the model. We use such a metric at inference time too, however, differently from how we define it in \eqref{eq:nmse_obj}, at test time we evaluate the NMSE on each data sample separately
to plot the CDF.

\subsection{DLISTA versus ISTA with dictionary learning}

\begin{figure}[t]
    \centering
    \includegraphics[width=\linewidth]{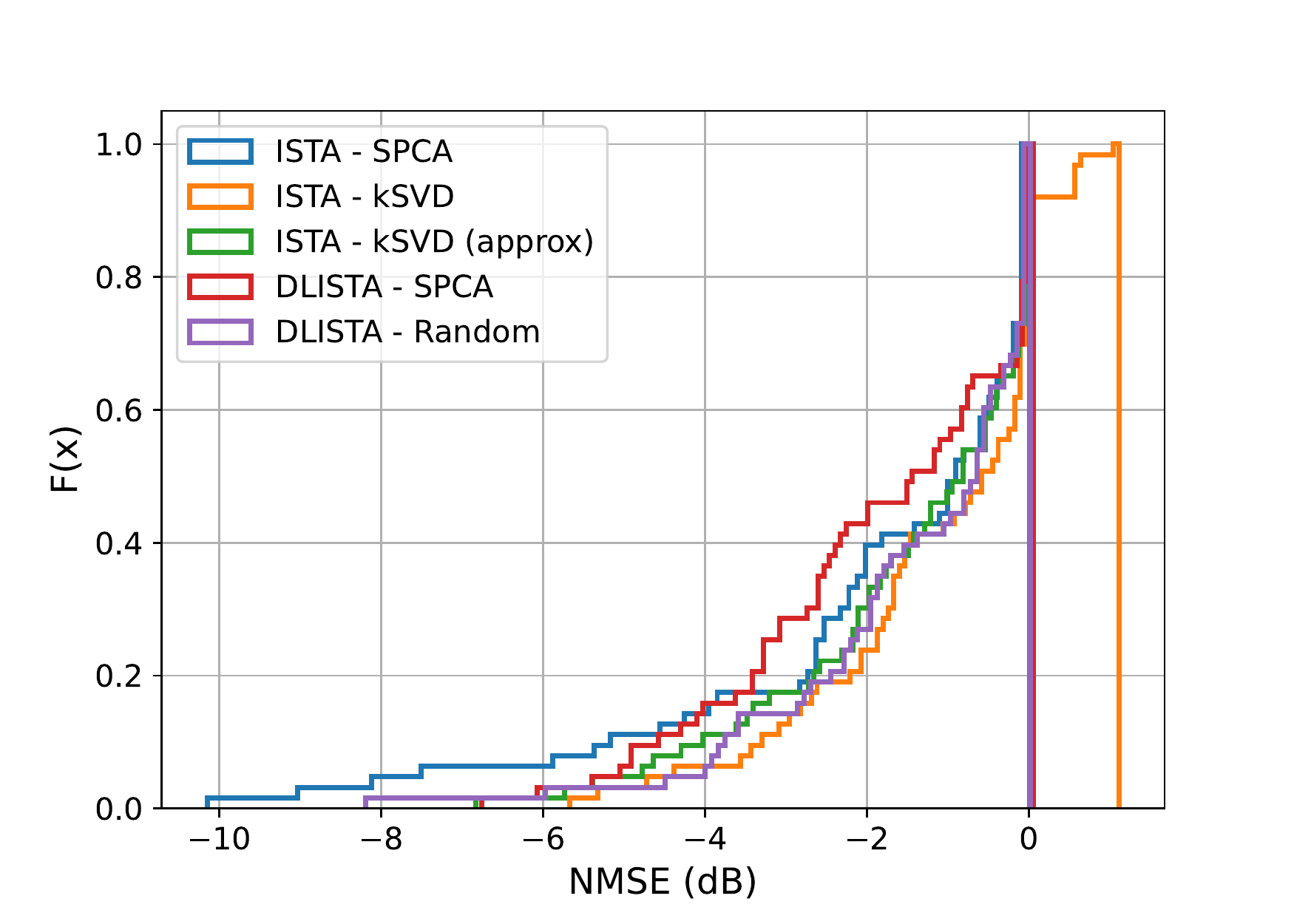}
    \caption{
    \ac{ISTA} and \modelacro{} models with different initialization of the $\Psi$s matrices.
    We consider 10 iterations, 200 atoms and non-shared weights, with exception of \modelacro{} with a randomly initialized dictionary and 2000 atoms.}
    \label{fig:ista_dlista}
\vspace{-4mm}
\end{figure}

As mentioned in~\secref{sec:dict_learning}, we formulate our model as an unrolled parametrized reformulation of \ac{ISTA}~\cite{blumensath_iterative_2009}. For such a reason, we use it as a baseline. In order to select the best hyperparameters for \ac{ISTA}, we adopt an extensive grid search on the step size and the threshold values. Concerning the dictionary for \ac{ISTA}, we employ sparse PCA and kSVD (since \ac{ISTA} has no learnable parameters, we could not use randomly initialized dictionaries). We report in~\Figref{fig:ista_dlista} a performance comparison among the various \ac{ISTA} and \modelacro{} models for different type of dictionaries. In~\Figref{fig:ista_dlista} we report the performance from the best models only. Interestingly, we did not notice any clear advantage in using kSVD over sparse SPCA for \ac{ISTA}. Thus, being the former algorithm more computationally demanding than the latter algorithm, we decide to initialize the $\mPsi$'s for \modelacro{} using sparse PCA and random sampling only. We empirically found that the best results from all the models correspond to configurations with non-shared weights and 10 iterations. Note that, DLISTA-SPCA provides the best result for 80\% quantile, although ISTA-SPCA achieves better results for $\sim$10\% quantile. As we will see in the next section, DLISTA performs much better in terms of spectral efficiency compared to ISTA despite comparable NMSE performance.
\begin{figure}
    \centering
    \includegraphics[width=\linewidth]{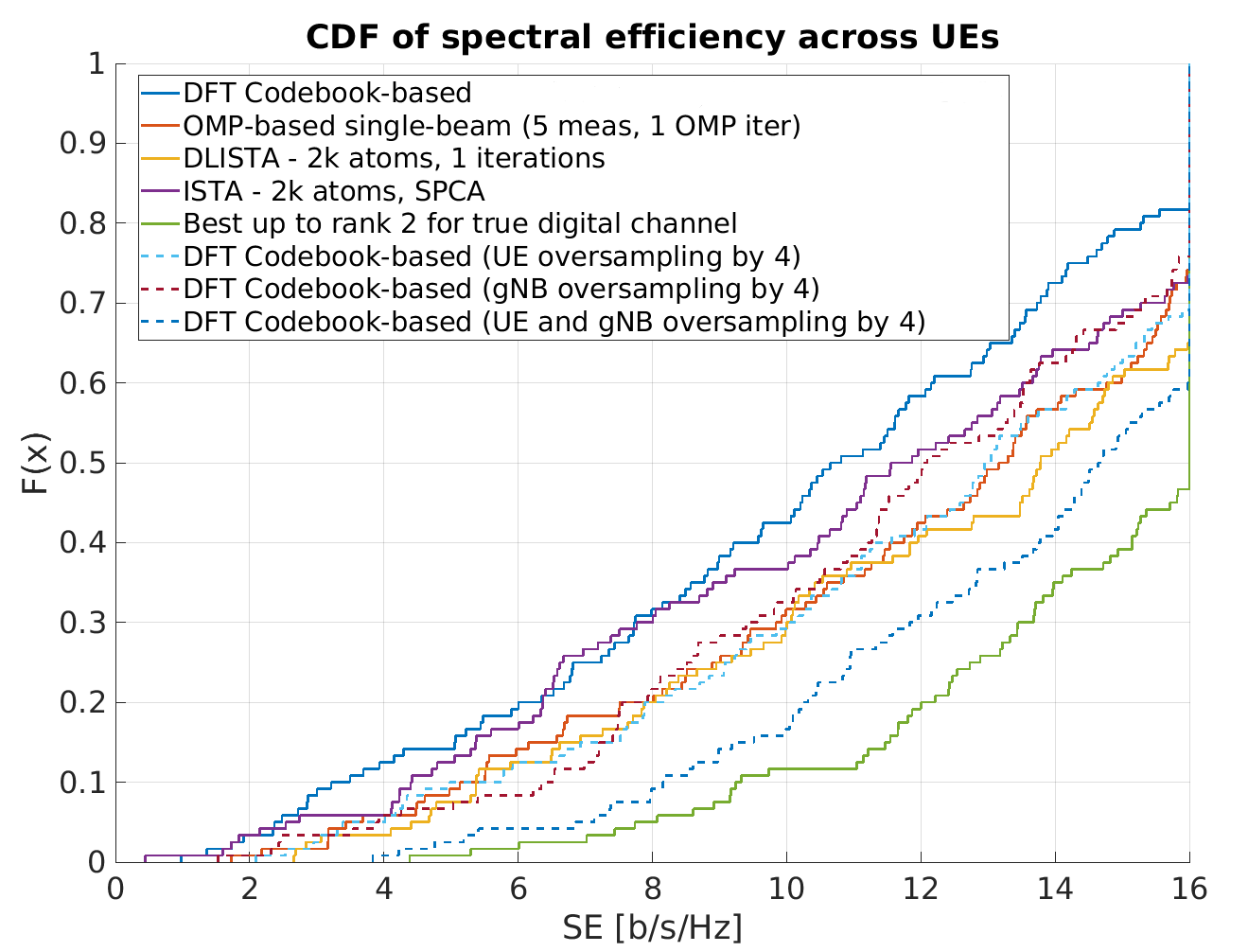}
    \caption{Spectral Efficiency CDF for 120 UEs in indoor hotspot deployment}
    \label{fig:SE_CDF_OMP_InH_5meas}
\vspace{-5mm}
\end{figure}
\vspace{-3mm}
\subsection{Spectral Efficiency Results}
Based on the estimated channels, we customize the analog beams and compute the spectral efficiency. As mentioned before, OMP provides angular information, and the beams at UE and gNB are directed in such a way to be aligned with the AoA/AoD of the strongest cluster. For DLISTA, we sweep over an oversampled codebook and select the best one. As a first baseline, we consider best rank-2 performance for true digital channel. This provides an upper bound on what can be done with estimated raw channel, using rank-2 transmissions. To see how far the performance can be pushed with pre-defined codebook beams, we consider normal and oversampled beamforming codebooks on gNB and/or UE side. As it can be seen from the results in Figure \ref{fig:SE_CDF_OMP_InH_5meas}, the methods based on tailoring the transmit and receive beams to the estimated raw channel can lead to better spectral efficiency compared to methods based on pre-defined codebooks.  Note that DLISTA and OMP achieve comparable spectral efficiency performance, although ISTA based methods with SPCA underperforms. We make a final remark on the complexity of OMP versus DLISTA. In terms of number of iterations, OMP and DLISTA were comparable, partially because the indoor hotspot channel  had a few dominant clusters. In terms of dictionary dimension, however, DLISTA provided considerable boost by reducing the dimension to 2000 atoms from the original 524k atoms. Upon further investigations, the dictionary dimension for OMP and DLISTA could be further reduced to 32k and 200 respectively, still leaving a considerable gap.

\section{Conclusions}
We have introduced compressed sensing and machine learning based approach for estimating raw channel from analog beamformed measurements and used it for better beam selection. It was shown that beam design using knowledge of raw channel provides gains beyond methods relying on pre-defined codebooks. Besides, DLISTA provides an approach for channel estimation with low complexity. 

\bibliographystyle{IEEEtran}
\bibliography{references.bib}

\begin{thebibliography}{10}
\providecommand{\url}[1]{#1}
\csname url@samestyle\endcsname
\providecommand{\newblock}{\relax}
\providecommand{\bibinfo}[2]{#2}
\providecommand{\BIBentrySTDinterwordspacing}{\spaceskip=0pt\relax}
\providecommand{\BIBentryALTinterwordstretchfactor}{4}
\providecommand{\BIBentryALTinterwordspacing}{\spaceskip=\fontdimen2\font plus
\BIBentryALTinterwordstretchfactor\fontdimen3\font minus
  \fontdimen4\font\relax}
\providecommand{\BIBforeignlanguage}[2]{{%
\expandafter\ifx\csname l@#1\endcsname\relax
\typeout{** WARNING: IEEEtran.bst: No hyphenation pattern has been}%
\typeout{** loaded for the language `#1'. Using the pattern for}%
\typeout{** the default language instead.}%
\else
\language=\csname l@#1\endcsname
\fi
#2}}
\providecommand{\BIBdecl}{\relax}
\BIBdecl

\bibitem{rodriguez-fernandez_frequency-domain_2018}
J.~Rodríguez-Fernández, N.~González-Prelcic, K.~Venugopal, and R.~W. Heath,
  ``Frequency-{Domain} {Compressive} {Channel} {Estimation} for
  {Frequency}-{Selective} {Hybrid} {Millimeter} {Wave} {MIMO} {Systems},''
  \emph{IEEE Transactions on Wireless Communications}, vol.~17, no.~5, pp.
  2946--2960, May 2018.

\bibitem{venugopal_channel_2017}
K.~Venugopal, A.~Alkhateeb, N.~González~Prelcic, and R.~W. Heath, ``Channel
  {Estimation} for {Hybrid} {Architecture}-{Based} {Wideband} {Millimeter}
  {Wave} {Systems},'' \emph{IEEE Journal on Selected Areas in Communications},
  vol.~35, no.~9, pp. 1996--2009, Sep. 2017.

\bibitem{wang_beam_2019}
B.~Wang, M.~Jian, F.~Gao, G.~Y. Li, and H.~Lin, ``Beam {Squint} and {Channel}
  {Estimation} for {Wideband} {mmWave} {Massive} {MIMO}-{OFDM} {Systems},''
  \emph{IEEE Transactions on Signal Processing}, vol.~67, no.~23, pp.
  5893--5908, Dec. 2019.

\bibitem{ma_high-resolution_2020}
W.~Ma, C.~Qi, and G.~Y. Li, ``High-{Resolution} {Channel} {Estimation} for
  {Frequency}-{Selective} {mmWave} {Massive} {MIMO} {Systems},'' \emph{IEEE
  Transactions on Wireless Communications}, vol.~19, no.~5, pp. 3517--3529, May
  2020.

\bibitem{pati_orthogonal_1993}
Y.~Pati, R.~Rezaiifar, and P.~Krishnaprasad, ``Orthogonal matching pursuit:
  recursive function approximation with applications to wavelet
  decomposition,'' in \emph{Proceedings of 27th {Asilomar} {Conference} on
  {Signals}, {Systems} and {Computers}}, Nov. 1993, pp. 40--44 vol.1.

\bibitem{davis_adaptive_1994}
G.~M. Davis, S.~G. Mallat, and Z.~Zhang, ``Adaptive time-frequency
  decompositions,'' \emph{Optical Engineering}, vol.~33, no.~7, pp. 2183--2191,
  Jul. 1994.

\bibitem{blumensath_iterative_2009}
T.~Blumensath and M.~E. Davies, ``\BIBforeignlanguage{en}{Iterative hard
  thresholding for compressed sensing},'' \emph{\BIBforeignlanguage{en}{Applied
  and Computational Harmonic Analysis}}, vol.~27, no.~3, pp. 265--274, Nov.
  2009.

\bibitem{daubechies2004iterative}
I.~Daubechies, M.~Defrise, and C.~De~Mol, ``An iterative thresholding algorithm
  for linear inverse problems with a sparsity constraint,''
  \emph{Communications on Pure and Applied Mathematics: A Journal Issued by the
  Courant Institute of Mathematical Sciences}, vol.~57, no.~11, pp. 1413--1457,
  2004.

\bibitem{donoho_message-passing_2009}
D.~L. Donoho, A.~Maleki, and A.~Montanari,
  ``\BIBforeignlanguage{en}{Message-passing algorithms for compressed
  sensing},'' \emph{\BIBforeignlanguage{en}{Proceedings of the National Academy
  of Sciences}}, vol. 106, no.~45, pp. 18\,914--18\,919, Nov. 2009.

\bibitem{behrens_neurally_2021}
F.~Behrens, J.~Sauder, and P.~Jung, ``Neurally {Augmented} \{{ALISTA}\},'' in
  \emph{International {Conference} on {Learning} {Representations}}, 2021.

\bibitem{gregor_learning_2010}
K.~Gregor and Y.~LeCun, ``Learning fast approximations of sparse coding,'' in
  \emph{27th {International} {Conference} on {Machine} {Learning}, {ICML}
  2010}, 2010.

\bibitem{liu_alista_2019}
J.~Liu, X.~Chen, Z.~Wang, and W.~Yin, ``{ALISTA}: {Analytic} {Weights} {Are}
  {As} {Good} {As} {Learned} {Weights} in {LISTA},'' in \emph{International
  {Conference} on {Learning} {Representations}}, 2019.

\bibitem{chen_hyperparameter_lista_2021}
X.~Chen, J.~Liu, Z.~Wang, and W.~Yin, ``Hyperparameter tuning is all you need
  for lista,'' \emph{Advances in Neural Information Processing Systems},
  vol.~34, 2021.

\bibitem{aberdam_ada_lista_2021}
A.~Aberdam, A.~Golts, and M.~Elad, ``Ada-{LISTA}: {Learned} {Solvers}
  {Adaptive} to {Varying} {Models},'' \emph{IEEE Transactions on Pattern
  Analysis and Machine Intelligence}, pp. 1--1, 2021.

\bibitem{schnoor_generalization_2022}
E.~Schnoor, A.~Behboodi, and H.~Rauhut, ``Generalization {Error} {Bounds} for
  {Iterative} {Recovery} {Algorithms} {Unfolded} as {Neural} {Networks},''
  \emph{arXiv:2112.04364}, Jan. 2022.

\bibitem{schniter_channel_2014}
P.~Schniter and A.~Sayeed, ``Channel estimation and precoder design for
  millimeter-wave communications: {The} sparse way,'' in \emph{2014 48th
  {Asilomar} {Conference} on {Signals}, {Systems} and {Computers}}, Nov. 2014,
  pp. 273--277.

\bibitem{ma_sparse_2013}
Z.~Ma, ``\BIBforeignlanguage{en}{Sparse principal component analysis and
  iterative thresholding},'' \emph{\BIBforeignlanguage{en}{The Annals of
  Statistics}}, vol.~41, no.~2, pp. 772--801, Apr. 2013.

\bibitem{aharon2006k}
M.~Aharon, M.~Elad, and A.~Bruckstein, ``K-svd: An algorithm for designing
  overcomplete dictionaries for sparse representation,'' \emph{IEEE
  Transactions on signal processing}, vol.~54, no.~11, pp. 4311--4322, 2006.

\bibitem{rubinstein2008efficient}
R.~Rubinstein, M.~Zibulevsky, and M.~Elad, ``Efficient implementation of the
  k-svd algorithm using batch orthogonal matching pursuit,'' Computer Science
  Department, Technion, Tech. Rep., 2008.

\bibitem{3gpp_38901_2020}
{ETSI 3rd Generation Partnership Project (3GPP)}, ``Study on channel model for
  frequencies from 0.5 to 100 ghz (version 16.1.0),'' \emph{3GPP TR 38.901,
  Jun, Tech. Rep.}, 2020.

\end{thebibliography}
\end{document}